\documentclass[%
 reprint,
nofootinbib,
 amsmath,amssymb,
 aps,
prd,
]{revtex4-2}

\usepackage{color}
\usepackage{ulem}
\usepackage{slashed}

\newcommand{\tr}{{\rm Tr}\,}

\newcommand{\pt}{\partial}
\newcommand{\mub}{\hat{\mu}_B}
\newcommand{\ctld}{\chi}
\newcommand{\dmu}{\frac{\pt}{\pt\mub}}
\newcommand{\lbar}{\bar{\ell}}

\newcommand{\vx}{\vec{x}}

\usepackage{multirow}
\usepackage{tabularx}
\usepackage{booktabs}
\usepackage{graphicx}
\usepackage{makecell}
\usepackage{adjustbox}
\usepackage{dcolumn}
\usepackage{bm}
\usepackage{hyperref}

\begin{document}

\title{Scaling properties of net-baryon number fluctuations at the deconfinement critical point}

\author{Micha\l{}  Szyma\'nski}
\email{michal.szymanski@uwr.edu.pl}
\affiliation{Institute of Theoretical Physics, University of Wroc\l aw, plac  Maksa  Borna  9, PL-50204 Wroc\l aw, Poland}

\author{Pok Man Lo}
\affiliation{Institute of Theoretical Physics, University of Wroc\l aw, plac  Maksa  Borna  9, PL-50204 Wroc\l aw, Poland}

\author{Krzysztof Redlich}
\affiliation{Institute of Theoretical Physics, University of Wroc\l aw, plac  Maksa Borna  9, PL-50204 Wroc\l aw, Poland}
\affiliation{Polish Academy of Sciences PAN, Podwale 75, PL-50449 Wroc\l aw, Poland}

\author{Chihiro Sasaki}
\affiliation{Institute of Theoretical Physics, University of Wroc\l aw, plac  Maksa  Borna  9, PL-50204 Wroc\l aw, Poland}
\affiliation{International Institute for Sustainability with Knotted Chiral Meta Matter (WPI-SKCM$^2$), Hiroshima University, Higashi-Hiroshima, Hiroshima 739-8531, Japan}

\date{\today}

\begin{abstract}
We investigate the critical behavior of the first four cumulants of the net-baryon number near the deconfinement critical point in QCD in the limit of heavy quarks. By connecting baryon-number fluctuations to Polyakov loop susceptibilities, we analyze their mean-field scaling properties at zero and non-zero baryon chemical potentials. In the mean-field approximation we derive critical exponents  via the Landau theory and validate them through explicit numerical calculations in an effective Polyakov loop model. 
Using a Ginzburg–Landau criterion as a diagnostic of beyond–mean-field effects, we estimate the size of the critical region and find that it shrinks with increasing baryon density. By utilizing the exact scaling function in the  3D Ising
model, we estimate critical exponents beyond the mean-field approximation.

\end{abstract}

\maketitle

\section{Introduction}

Quantum chromodynamics (QCD) is expected to have a rich phase structure, determination of which is an ongoing experimental~\cite{Braun-Munzinger:2015hba,Luo:2017faz,Andronic:2017pug,Bzdak:2019pkr} as well as theoretical effort
~\cite{Fukushima:2010bq,Fukushima:2013rx,Guenther:2020jwe,Ratti:2022qgf,Aarts:2023vsf}. Besides physically relevant parameters, such as temperature and baryon chemical potential, or the magnetic field~\cite{Andersen:2014xxa,Miransky:2015ava,Endrodi:2024cqn,Adhikari:2024bfa}, the phase diagram is often investigated in the range of parameters which cannot be modified in the physical setting, such as quark mass and number of colors or flavors. While not directly relevant to the experiment, such studies are useful for understanding particular features of QCD. One possibility is to investigate QCD with larger than physical quark masses. Such choice is particularly suited for studying deconfinement because the effects due to the chiral physics become less prominent.

In the limit of infinite quark mass, deconfinement is a first-order phase transition~\cite{Boyd:1996bx} which can related to the spontaneous breaking of the center symmetry~\cite{Svetitsky:1982gs,Greensite:2003bk} with the Polyakov loop~\cite{Polyakov:1978vu,tHooft:1977nqb,Svetitsky:1982gs} serving as an order parameter. While the center symmetry is broken explicitly by dynamical quarks, the transition remains first-order unless the critical quark mass value is reached, at which the deconfinement becomes a second order phase transition~\cite{Green:1983sd}. This marks the deconfinement critical point, properties of which have been investigated using lattice simulations~\cite{Ejiri:2019csa,Cuteri:2020yke,Ashikawa:2024njc}, functional methods~\cite{Fischer:2014vxa} as well as effective models~\cite{Kashiwa:2012wa,Lo:2014vba,Lo:2020ptj,Pham:2021ftz,MariSurkau:2025dfo,Konrad:2025ndq}.

When calculated on the lattice, the Polyakov loop suffers from the renormalization scheme dependence~\cite{Bazavov:2016uvm,Weber:2017dmi} which calls into question validity of physical results extracted from this observable. Thus, it is important to investigate also other quantities sensitive to deconfinement, such that a consistent picture can be obtained. One of possibilities is to investigate susceptibilities of the baryon number. On the lattice, these quantities have been extensively studied for physical quark masses~\cite{Borsanyi:2013hza,Bazavov:2013dta,Bazavov:2020bjn,Borsanyi:2023wno} and are the cornerstone of experimental searches of the QCD critical point~\cite{Luo:2017faz,Asakawa:2015ybt,Bzdak:2019pkr}. Recently, properties of these quantities in the heavy quark mass limit at the first-order phase transition were examined theoretically using hopping parameter expansion~\cite{Tohme:2025nzw}. It has been found that these quantities should have a discontinuity at the critical temperature. This suggests that {susceptibilities of the
baryon number} are indeed useful for studying QCD in the heavy-quark limit. 

The aim of this work is to explore utility of net-baryon number susceptibilities for investigating properties of the deconfinement critical point. For physical quark masses these quantities are expected to diverge at the critical point with the strength of divergence increasing with the order of the susceptibility~\cite{Stephanov:2008qz}. We investigate the critical behavior of the first four cumulants of the net-baryon number in the proximity of the deconfinement critical point at zero and non-zero net-baryon densities.  In the mean-field (MF) approximation, these quantities are calculated using an effective Polyakov loop model. To investigate their MF criticality, we connect the fluctuations of the net-baryon number with the susceptibilities of Polyakov loop. Using the Landau-type potential we extract the MF critical exponents and confirm our prediction with the explicit numerical calculation. We also investigate critical exponents beyond the MF approximation using the scaling function approach.

While critical exponents are determined by the universality class of the system, the size of the critical region (inside which the MF approximation breaks down) is non-universal. Reliable estimates of this region are relevant for phenomenology. One possible prescription is to investigate the contours of the net-baryon number susceptibilities relative to the free susceptibilities around the critical point~\cite{Schaefer:2006ds,Schaefer:2011ex}. Here, we estimate the size of the critical region using the Ginzburg criterion~\cite{Goldenfeld:1992qy,Kopietz:2010zz} (see also Ref.~\cite{Hatta:2002sj} for the discussion regarding the critical region of QCD). In Ref. \cite{Lo:2026xuc} the MF formulation of this criterion has been discussed. Following that work we formulate the Ginzburg criterion for our model and discuss the size of the critical region. Particularly, we find that it shrinks with increasing baryon density. 

The paper is organized as follows: in Sec.~\ref{sec:model} we present details of the model used in this work. Section~\ref{sec:results} shows numerical results for net-baryon number fluctuations at zero and non-zero baryon chemical potential. In Sec.~\ref{sec:scaling} we discuss the critical behavior near the deconfinement critical point: Sec.~\ref{sec:MF} contains analytic predictions on MF critical exponents, in Sec:~\ref{sec:num_scaling} we confirm these predictions numerically and in Sec.~\ref{sec:beyond_mf} we us the scaling function approach to obtain the prediction on the critical behavior of net-baryon number susceptibilities in the $Z_2$ universality class. In Sec.~\ref{sec:crit_region} the estimate of the critical region based on the Ginzburg criterion is discussed and in Sec.~\ref{sec:conclusions} we present our conclusions. 

\section{Effective Polyakov loop model}
\label{sec:model}
To model thermodynamics of QCD with heavy quarks we use an effective Polyakov loop potential~\cite{Kashiwa:2012wa,Lo:2014vba},
\begin{equation}
    U(\ell,\lbar)=U_{G}(\ell,\lbar)+U_Q(\ell,\lbar)\,,
\label{eq:pot_gen}
\end{equation}
where $\ell$, $\lbar$ are the traced Polyakov loop and its conjugate~\cite{Polyakov:1978vu, McLerran:1981pb},
\begin{equation}
    \ell(\vx)=\frac{1}{3}\tr_Ce^{ig\int_0^\beta A_4(\tau,\vx)d\tau}\,,
\end{equation}
where $\tr_C$ stands for the trace over color and $\beta=1/T$. $U_{G}$ is the pure gauge part and $U_Q$ contains the contribution due to the dynamical quarks. For the former, we choose the potential obtained in Ref.~\cite{Lo:2013hla},
\begin{eqnarray}
    \frac{U_G(\ell,\bar{\ell})}{T^4}=&&-\frac{1}{2}A(t)\ell\lbar+B(t)\ln M_H(\ell,\lbar)\nonumber\\
    &&+\frac{1}{2}C(t)(\ell^3+\lbar^3)+D(t)(\ell\lbar)^2\label{eq:gluon_potential}\,,    
\end{eqnarray}
where $t=T/T_d$ with $T_d=270\,$MeV being the deconfinement temperature of the pure SU(3) theory, and
\begin{equation}
    M_H(\ell,\bar{\ell})=1-6\ell\bar{\ell}+4(\ell^3+\lbar^3)-3(\ell\lbar)^2
\end{equation}
is the SU(3) Haar measure~\cite{Fukushima:2017csk}. Coefficients of this potential were determined using the LQCD data on the pure SU(3) equation of state, Polyakov loop expectation value and its susceptibilities (see Ref.~\cite{Lo:2013hla} for their functional form and further discussion). 

To model the quark contribution we employ the one-loop fermion determinant in the $A_4$ background which in terms of Polyakov loop and its conjugate reads~\cite{Kashiwa:2012wa,Lo:2014vba}
\begin{widetext}
\begin{eqnarray}
    U_Q(M,\ell,\bar{\ell})=-2TN_f\int\frac{d^3q}{(2\pi)^3}\big[&&\ln(1+3\ell e^{-\beta(E-\mu)}+3\bar{\ell}e^{-2\beta(E-\mu)}+e^{-3\beta(E-\mu)})\nonumber\\
    +&&\ln(1+3\bar{\ell} e^{-\beta(E+\mu)}+3\ell e^{-2\beta(E+\mu)}+e^{-3\beta(E+\mu)}) \big]\label{eq:quark_potential}\,, 
\end{eqnarray}
\end{widetext}
where $E=\sqrt{p^2+M^2}$. 

In the following, we neglect the contribution due to the quark-quark interaction and thus the quark mass remains constant. In fact, we checked the effect due to the quark dynamics by considering the Nambu--Jona-Lasinio (NJL)-type interaction~\cite{Nambu:1961tp,Nambu:1961fr} (see also Refs.~\cite{Klevansky:1992qe,Buballa:2003qv} for the review),
\begin{equation}
    \mathcal{L}_{\text{quark-quark}}=G\left[(\bar{q}q)^2+(\bar{q}i\gamma_5\vec{\tau}q)^2\right]\,,
\end{equation}
and found that the net-baryon number fluctuations near the deconfinement critical point show the same qualitative behavior as when this contribution is neglected. We also note that the commonly used 3-dimensional momentum cutoff is not suited for extension of the (P)NJL model to heavy quarks -- this cutoff suffers from the positive chiral condensate problem at finite temperature~\cite{Ruivo:2011fg,Pasqualotto:2023hho} which becomes severe for larger bare quark masses, effectively yielding the dressed quark mass lighter than the bare one. The non-local extension of the NJL model (see for example Ref.~\cite{Sasaki:2006ww}) is, however, free of such issue.

In this work, we consider the MF approximation in which the expectation values of the Polyakov loop and its conjugate are determined from the gap equations,
\begin{eqnarray}
\frac{\pt U}{\pt\ell} =0\,,~~~
\frac{\pt U}{\pt\lbar} =0
\label{eq:gap_eqs}
\end{eqnarray}
 At finite density, the expectation values of $\ell$ and $\lbar$ are both real, but different which is the consequence of the complex nature of the fermion determinant~\cite{Sasaki:2006ww,Fukushima:2017csk}. Furthermore, fluctuations of the order parameter are linked to the inverse curvature of the potential,
\begin{equation}
    \chi=\mathcal{C}^{-1}
\end{equation}
where $\mathcal{C}$ is the curvature matrix,
\begin{eqnarray}
    \mathcal{C}_{ij}&=&\frac{\partial^2 U/T^4}{\partial\phi_i\partial\phi_j}
    \label{eq:chi_def}
\end{eqnarray}
with $\phi_1\equiv \ell$, $\phi_2\equiv \lbar$. These susceptibilities can be related to susceptibilities of the real ($\chi_L$) and imaginary parts ($\chi_T$) of the Polyakov loop~\cite{Lo:2014vba},
\begin{eqnarray}
    \chi_{T}&=&\frac{1}{2}\bigg[\chi_{\ell\lbar}-\frac{1}{2}(\chi_{\ell\ell}+\chi_{\lbar\lbar})\bigg]\,,\nonumber\\
    \chi_{L}&=&\frac{1}{2}\bigg[\chi_{\ell\lbar}+\frac{1}{2}(\chi_{\ell\ell}+\chi_{\lbar\lbar})\bigg]\,.
    \label{eq:chi_T_L}
\end{eqnarray}
We refer the reader to Refs.~\cite{Lo:2014vba,Szymanski:2020stb} in which the expectation value of Polyakov loop and its conjugate as well as their susceptibilities near the deconfinement critical point were investigated for the effective Polyakov loop potential~\eqref{eq:gluon_potential}. Particularly, it has been observed that at $\mu_B=0$ the real susceptibility diverges while the imaginary one stays finite.

The aim of this work is to examine cumulants of the net-baryon number, defined as
\begin{equation}
\chi_n^B=\frac{\partial^n p}{\partial (\mub)^n}\bigg\vert_{T=const.}\,,
\end{equation}
where $p=P/T^4$ and $\mub=\mu_B/T$. In the current approach the pressure is linked to the effective potential via $P=-U(T,\mu,\ell(T,\mu),\bar{\ell}(T,\mu))$ with $\ell(T,\mu)$ and $\lbar(T,\mu)$ being the solutions of the gap equations, and thus the $\mu_B$-dependence of MFs has to be taken into account when performing the derivatives. Although $\chi_n^B$ can be computed numerically, it may be useful to express these cumulants in terms of Polyakov loop susceptibilities~\eqref{eq:chi_def}. Following the strategy outlined in Ref.~\cite{Fukushima:2017csk}, we obtained explicit expressions for $\chi_n^B$, $n=1,..,4$ (see Appendix~\ref{sec:app_chiB} for summary of the results). As an example, we apply these formulas to investigate the MF critical exponents of net-baryon number susceptibilities near the deconfinement CP (see Sec.~\ref{sec:scaling}).

\section{Net-baryon number susceptibilities}
\label{sec:results}

\begin{figure*}
    \centering
    \includegraphics[width=0.49\linewidth]{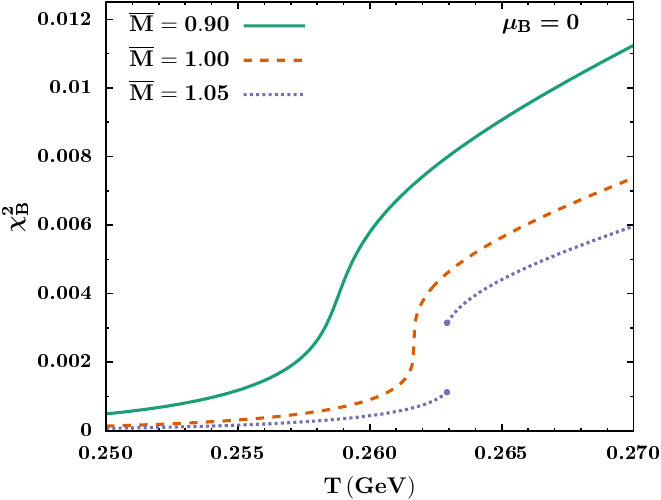}
    \includegraphics[width=0.49\linewidth]{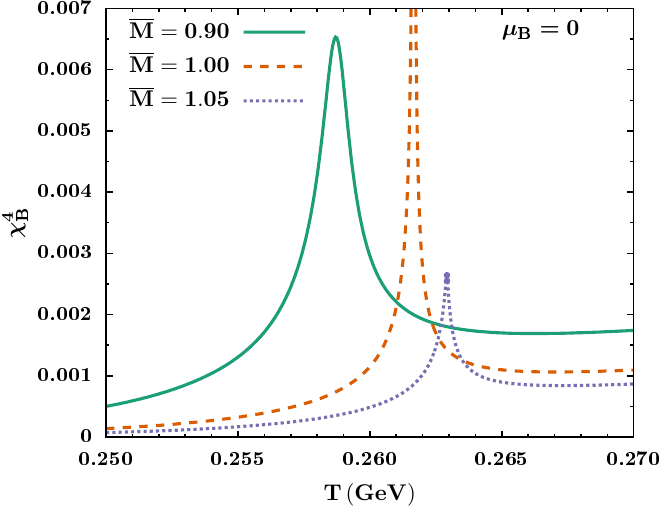}
    \caption{$\chi_2^B$ (left) and $\chi_4^B$ (right) in function of temperature for the vanishing chemical potential in case of the crossover ($\overline{M}=0.9$, green solid lines), critical point ($\overline{M}=1$, orange dashed lines), and first-order phase transition ($\overline{M}=1.05$, dotted violet line).}
    \label{fig:chi_B_mu_0}
\end{figure*}

\begin{figure}
    \centering
    \includegraphics[width=\linewidth]{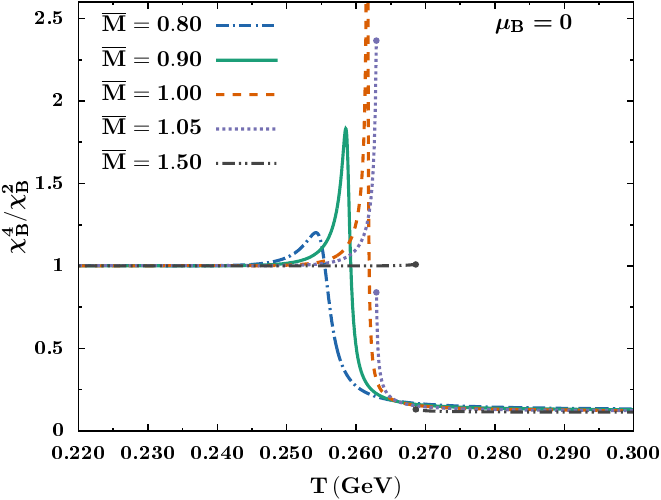}
    \caption{The $\chi_4^B/\chi_2^B$ in function of temperature at the vanishing chemical potential for $\overline{M}=0.8$ (blue dash-dotted line), $0.9$ (green solid line), $1.0$ (orange dashed line), $1.05$ (violet dotted line) and $1.5$ (black double-dot dashed line).}
    \label{fig:kurtosis}
\end{figure}

\begin{figure*}
    \centering
    \includegraphics[width=0.49\linewidth]{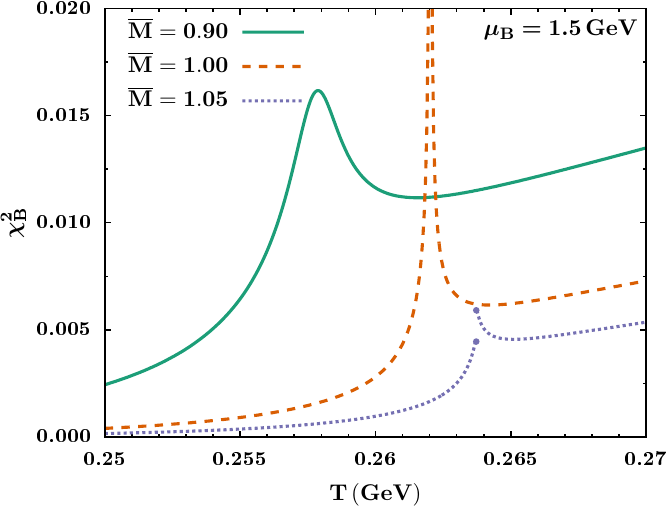}
    \includegraphics[width=0.49\linewidth]{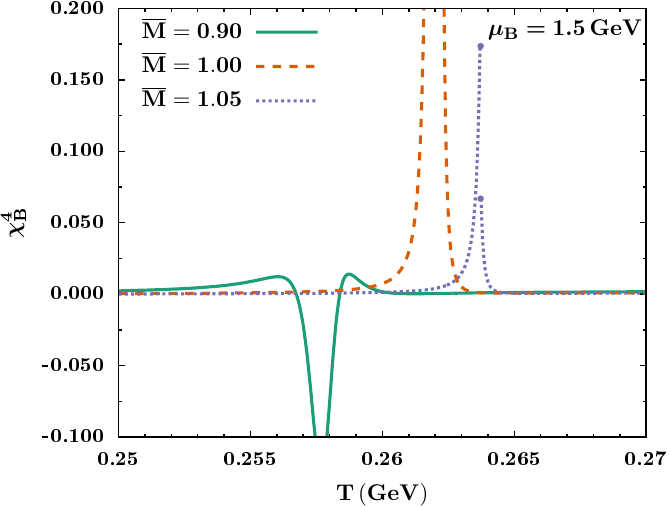}
    \caption{The same as in Fig.~\ref{fig:chi_B_mu_0} but for $\mu_B=1.5\,$GeV with $\overline{M}=M/M_{CP}^{\mu_B=1.5\,\text{GeV}}$.}
    \label{fig:chi_B_mu_500}
\end{figure*}

In this section we present numerical results on net-baryon number fluctuations near the deconfinement critical point. We consider $N_f=2$ case with degenerate quark masses. For $\mu_B=0$, we find the critical quark mass to be $M_{CP}^{\mu_B=0}=1.35160\,$GeV and critical temperature $T_{CP}^{\mu_B=0}=261.67\,$MeV (see Appendix \ref{sec:CP_determination} for details of determination of the CP location) which gives $M_{CP}^{\mu_B=0}/T_{CP}^{\mu_B=0}\approx 5.16$. Assuming that in the limit of heavy quarks the mass of the lightest pseudoscalar meson can be approximated by $2$ times the quark mass, we find this value to be lower than recent LQCD estimations, which find $M_{PS}/T_c \approx 18$~\cite{Cuteri:2020yke,Ashikawa:2024njc} (these results were not continuum extrapolated). The values of the critical temperature and quark mass depend also sensitively on the used Polyakov loop potential~\cite{Kashiwa:2012wa}.

The corresponding results for  $\chi_2^B$ (left) and $\chi_4^B$ (right) for different values of $\overline{M}= M/M_{CP}^{\mu_B=0}$ can be seen in Fig.~\ref{fig:chi_B_mu_0}. Green solid lines correspond to the crossover with $\overline{M}=0.9$, orange dashed ones to $\overline{M}=1$ and violet dotted lines $\overline{M}=1.05$, the first-order phase transition.  We observe that $\chi_2^B$ becomes increasingly steeper as the quark mass tends towards the critical value, but it does not diverge at the critical point. For masses larger than the critical value, the system undergoes a first-order phase transition and, in consequence, $\chi_2^B$ is discontinuous. On the other hand, the fourth order cumulant has a clear peak structure at the crossover and diverges at the critical point. As in the previous case, it is discontinuous at the first-order phase transition. We also observe that both quantities become suppressed with the increasing quark mass. 

In Fig.~\ref{fig:kurtosis} we show the kurtosis of the net-baryon number, $\chi^B_4/\chi_2^B$, calculated for $\overline{M}=0.8$ (blue dash-dotted line), $0.9$ (green solid line), $1.0$ (orange dashed line), $1.05$ (violet dotted line) and $1.5$ (black double-dot dashed line). Far from the (pseudo)critical temperature the $\chi^B_4/\chi_2^B$ ratio stays nearly constant, and is approximately equal to $1$ in the confined and  $1/9$ in the deconfined phase. The sharp peak develops for $T\approx T_c$, which follows from the critical behavior of $\chi^B_4$. As the quark mass exceeds the critical value, the net-baryon number kurtosis approaches a step-function behavior which indicates that it may be an excellent probe of deconfinement also in the heavy-quark limit (such behavior was discussed in Ref.~\cite{Fukushima:2017csk}, see also Ref.~\cite{Tohme:2025nzw} for a recent lattice QCD discussion).

Non-zero baryon chemical potential enhances the explicit center symmetry breaking~\cite{Lo:2014vba} and thus the critical value of the quark mass should be larger than the corresponding $\mu_B=0$ value. Indeed, for $\mu_B=1.5\,$GeV we find $M_{CP}^{\mu_B=1.5\,\text{GeV}}=1.78911\,$GeV and $T_{CP}^{\mu_B=1.5\,\text{GeV}}=262.08\,$MeV. We observe that in this case the behavior of $\chi_2^B$ substantially changes, see the left panel of Fig.~\ref{fig:chi_B_mu_500} (with $\overline{M}=M/M_{CP}^{\mu_B=1.5\,\text{GeV}}$ and the color convention being the same as in Fig.~\ref{fig:chi_B_mu_0}): it has a prominent peak at the crossover and becomes divergent at the CP . The fourth order cumulant (the right panel of Fig.~\ref{fig:chi_B_mu_500}) remains divergent at the CP but shows a more complicated structure than its zero-density counterpart, as it acquires a double-peak structure and may become negative (the green solid line). 

\begin{figure*}
    \centering
    \includegraphics[width=0.49\linewidth]{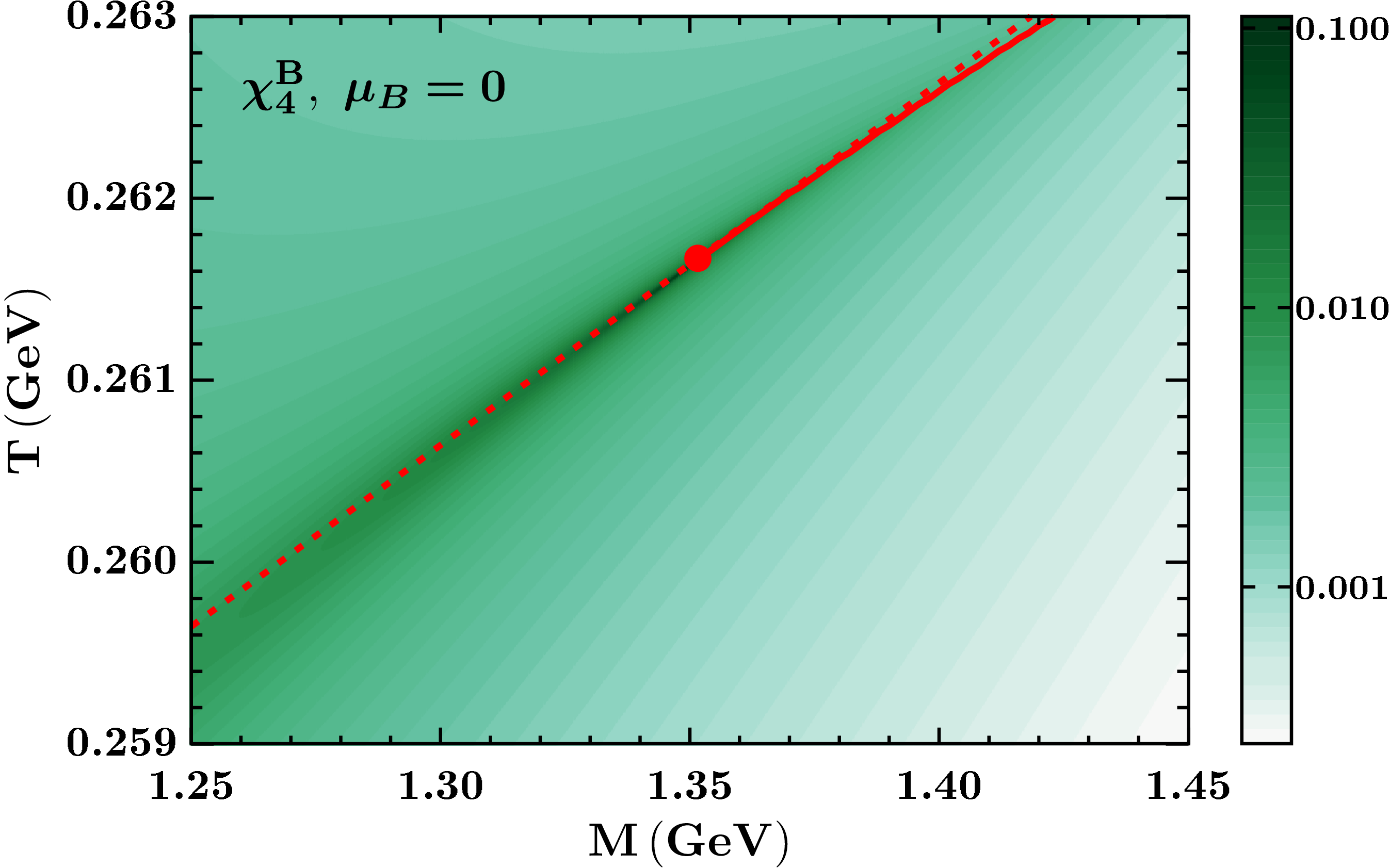}
    \includegraphics[width=0.49\linewidth]{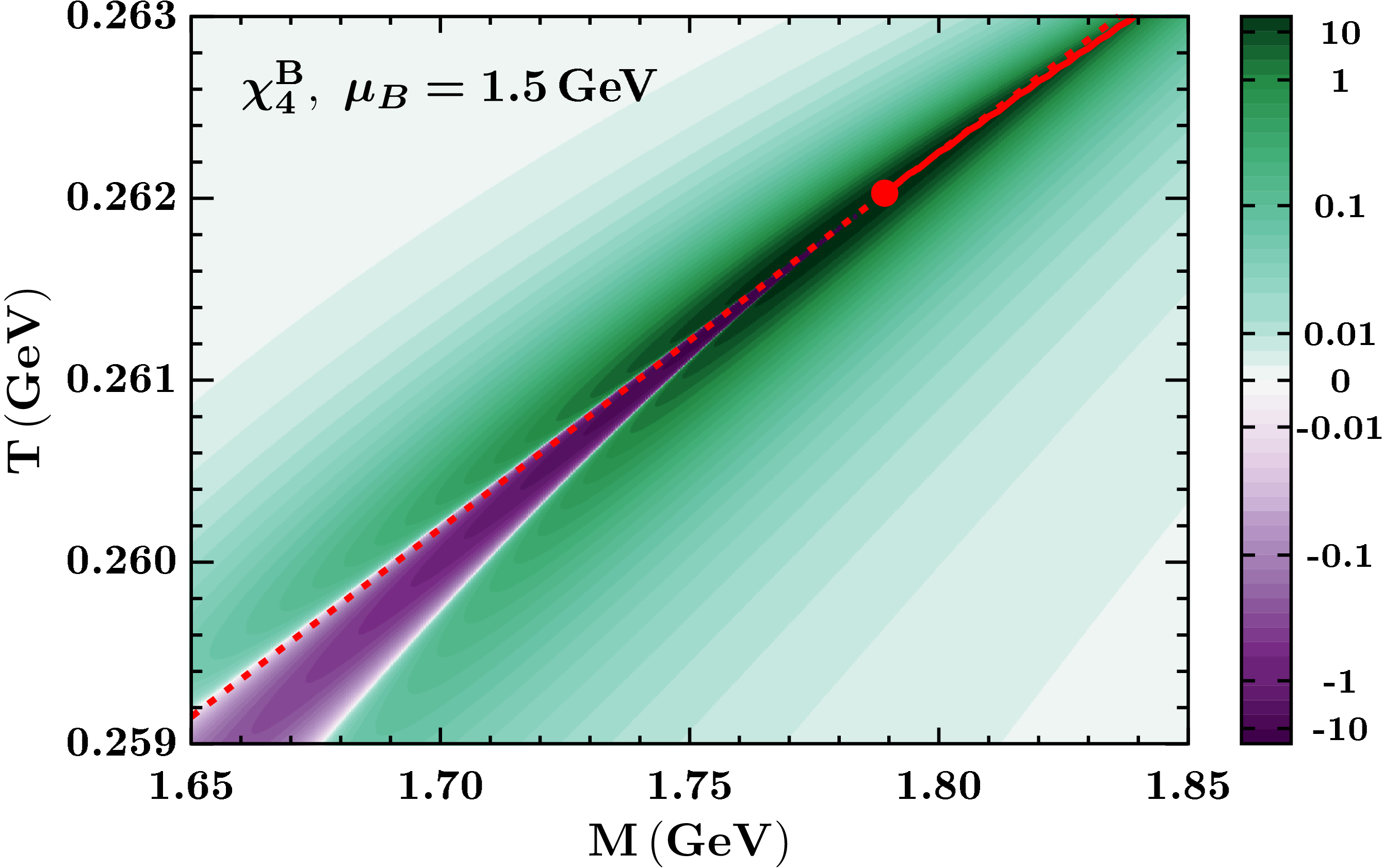}
    \caption{Fourth order baryon number cumulant in the proximity of the critical point in function of quark mass and temperature for $\mu_B=0$ (left) and $\mu_B=1.5\,$GeV (right). Solid red line -- first-order phase transition, dot -- deconfinement critical point, dashed line -- line tangent to the critical line at the critical point. Color intensity indicates the magnitude in the positive (green) and negative (purple) values.}
    \label{fig:chi4_3D}
\end{figure*}

A further insights into the structure of the fourth order cumulant at zero and non-zero densities can be obtained from Fig.~\ref{fig:chi4_3D} where a two-dimensional projection on $(M,\,T)$ plane is shown. Green color corresponds to positive values and purple to negative ones. The solid red line is the line of the first-order phase transition, the red dot marks the critical point and the red dashed line is the tangent to to the critical line at the CP (see the discussion in Sec.~\ref{sec:num_scaling}). At $\mu_B=0$ (left), $\chi_4^B$ is positive and gradually increases as the CP is approached. At finite density (right) the qualitative behavior of this quantity changes -- a region where it is negative develops. We also note that that the divergence is considerably stronger than in the previous case. 

\begin{figure}
    \centering
    \includegraphics[width=\linewidth]{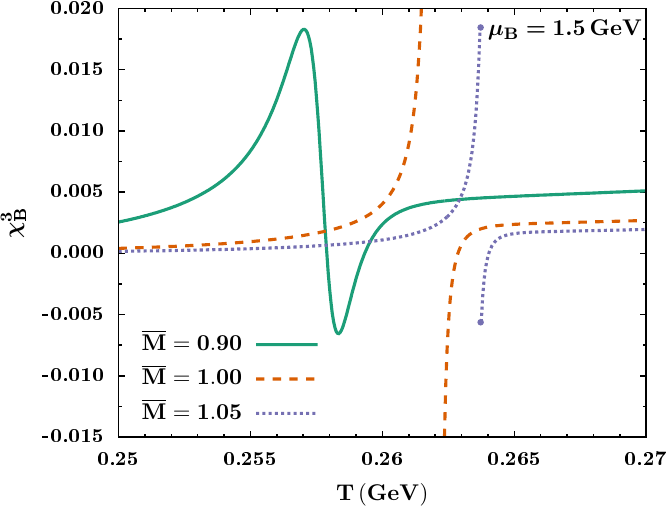}
    \caption{$\chi_3^B$ in function of temperature for $\mu_B=1.5\,$GeV with the same values of $\overline{M}$ as in Fig.~\ref{fig:chi_B_mu_500}.}
    \label{fig:chi3_vs_T}
\end{figure}

\begin{figure}
    \centering
    \includegraphics[width=\linewidth]{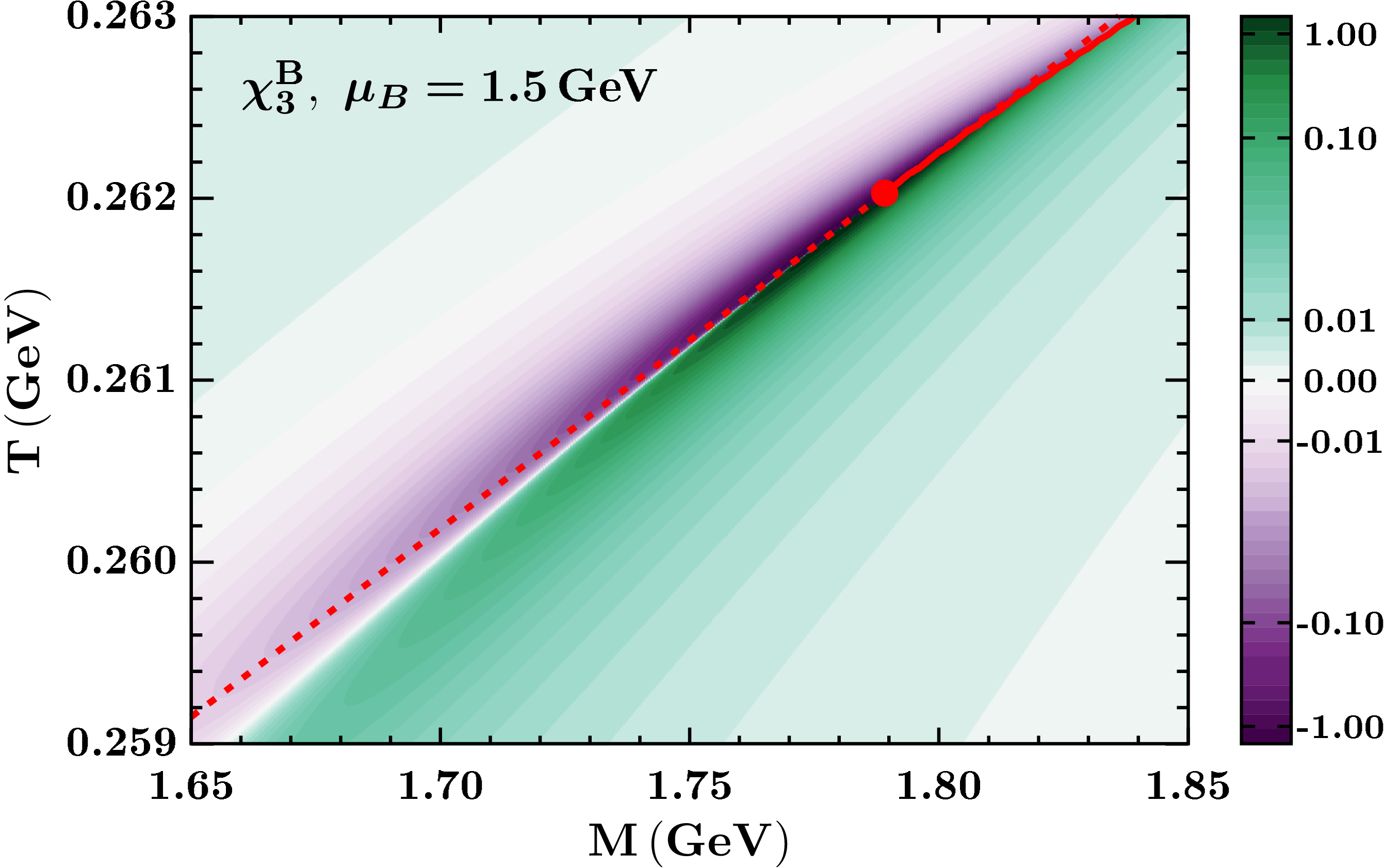}
    \caption{Third order baryon number susceptibility in the proximity of the critical point for $\mu_B=1.5\,$GeV. Solid red line -- first-order phase transition, dot -- deconfinement critical point, dashed line -- line tangent to the critical line at the critical point. Color intensity indicates the magnitude in the positive (green) and negative (purple) values.}
    \label{fig:chi3_3D}
\end{figure}

Additionally, at finite density the third order cumulant is non-zero (see Fig.~\ref{fig:chi3_vs_T}, with the color convention the same as in Fig \ref{fig:chi_B_mu_500}). We observe that it diverges to $\pm\infty$ depending on the direction of approach to CP (the orange dashed line). This is further confirmed from the two-dimensional projection of this quantity on the $(M,T)$ plane which shows two regions with the different signs of $\chi_B^3$, see Fig.~\ref{fig:chi3_3D}. 

\section{Critical behavior of net-baryon number cumulants}
\label{sec:scaling}
\subsection{MF critical exponents}
\label{sec:MF}
To investigate the MF criticality of the net-baryon number fluctuations, we consider a Landau-type potential which captures the MF criticality of the Ising model,
\begin{equation}
        \frac{U(\varphi,t,h)}{T^4}=u_0+\frac{1}{2}u_2 t\varphi^2+\frac{1}{4}u_4\varphi^4-h\varphi\,,
\label{eq:landau_potential}
\end{equation}
 where $u_2$ and $u_4$ are parameters of the potential and $t$ and $h$ are reduced temperature and magnetic field of the Ising model which are functions of the physical variables $T$, $\mu_B$ and $M$. We assume that close to CP they can be approximated as
\begin{eqnarray}
    t(T,\mu_B,M)&=&A \frac{T-T_c}{T_c}+B \frac{\mu_B^2}{T_c^2}+C 
    \frac{M-M_c}{M_c}\nonumber\\
    h(T,\mu_B,M)&=&D \frac{T-T_c}{T_c}+E \frac{\mu_B^2}{T_c^2}+F \frac{M-M_c}{M_c}\nonumber\\
    \label{eq:h_t_param}
\end{eqnarray}
where $A,$...,$F$ are coefficients which determine the direction of the approach to the CP (with $AF-CD\neq0$). The particle-antiparticle symmetry requires that $P(\mu_B)=P(-\mu_B)$ and thus the term linear in $\mu_B$ vanishes~\cite{Ejiri:2005wq}.

Furthermore, critical exponents depend on the direction from which the critical point is approached~\cite{Hatta:2002sj}. For the path which is asymptotically tangential to the critical line, susceptibility diverges as $\vert d\vert^{-\gamma}$ (where $d$ quantifies the distance to the CP), while for other directions the divergence is weaker, $\vert t\vert^{-\varepsilon}$, where $\varepsilon=\gamma/(\beta\delta)$ and $t=(T-T_c)/T_c$. Particularly, in the MF approximation $\gamma=1$, $\beta=1/2$ and $\delta=3$ and thus $\varepsilon=2/3$. Additionally in this approximation, the order parameter scales as $\vert d\vert^{\beta}$ for paths asymptotically tangential to the critical line and $\vert t\vert ^{1/\delta}$ for other directions.

To investigate the critical behavior of the net-baryon number cumulants, we apply Eqs.~\eqref{eq:chi2B_ex}, \eqref{eq:chi3B_ex} and \eqref{eq:chi4B_ex} to the potential~\eqref{eq:landau_potential}. At $\mu_B=0$ we find that the critical contribution to the second order cumulant is proportional to the order parameter and thus it stays finite at the critical point,
\begin{equation}
    \chi_2^B\vert_{crit.}\propto\varphi\,,
    \label{eq:chi2_mu_0}
\end{equation}
the third order cumulant vanishes and the critical contribution to the fourth order one is proportional to the order parameter susceptibility,
\begin{equation}
    \chi_4^B\vert_{crit.}\propto\chi\,.
    \label{eq:chi4_mu_0}
\end{equation}
At the finite chemical potential the critical behavior of net-baryon number cumulants is modified considerably. For the second-order cumulant we obtain
\begin{equation}
    \chi_2^B\vert_{crit.}\propto\chi\,.
    \label{eq:chi2_mu}
\end{equation}
For the third-order cumulant the leading singularity is
\begin{equation}
    \chi_3^B\vert_{crit.}\propto\varphi\chi^3
    \label{eq:chi3_mu}
\end{equation}
and thus for paths that are not asymptotically tangential to the critical line, the critical exponent is $3\gamma/(\beta\delta)-1/\delta=5/3$. For paths asymptotically tangential to the critical line, the direction of the approach has to be taken into account -- since the leading singularity is proportional to the order parameter, this term is non-zero only in the ordered phase ($t<0$) and vanishes in the disordered phase ($t>0$). For the former, the critical exponent is $3\gamma-\beta=2.5$ For the latter, the sub-leading term has to be considered,
\begin{equation}
    \chi_3^B\vert_{sub.}\propto\chi^2 \,
    \label{eq:chi3_mu_2}
\end{equation}
and the expected critical exponent is $2\gamma=2$. Finally, for the fourth order cumulant we find
\begin{equation}
    \chi_4^B\vert_{crit.}=c_1\chi^4+c_2\varphi^2\chi^5+...\,,
    \label{eq:chi4_mu}
\end{equation}
where $c_1$, $c_2$ are the corresponding coefficients and $...$ denote less singular terms. For the MF values of critical exponents of both terms diverge with the same rate -- the critical exponent of $\chi^B_4$ is $4$ for paths asymptotically tangential to the critical line and $8/3$ for other directions.

\subsection{Numerical determination of MF critical exponents}
\label{sec:num_scaling}

To test predictions from the previous section, we numerically investigate the net-baryon number cumulants near the critical point for $\mu_B=0$ and $1.5$\,GeV.  We first discuss a case of the path which is not tangential to the critical line. To this end, for a given chemical potential we set the quark mass to the corresponding critical value and approach the critical point along the temperature direction. To extract the critical exponents, we perform a linear fit
\begin{equation}
    \log_{10}\,\chi_n^B=-\varepsilon^-_{n}\,\log_{10}\,[(T_{CP}-T)/T_{CP}]+b^-
    \label{eq:linear_fit_Tm}
\end{equation}
for $T<T_{CP}$ and 
\begin{equation}
    \log_{10}\,\chi_n^B=-\varepsilon^+_{n}\,\log_{10}\,[(T-T_{CP})/T_{CP}]+b^+
    \label{eq:linear_fit_Tp}
\end{equation}
for $T>T_{CP}$. 

\begin{figure}
    \centering
    \includegraphics[width=\linewidth]{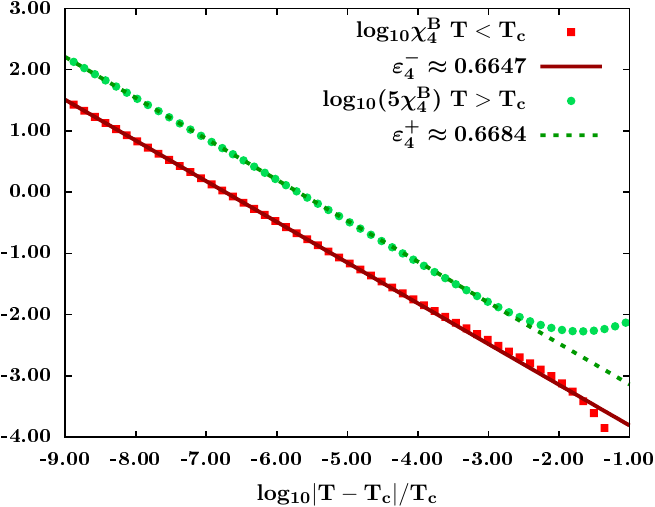}
    \caption{4th order susceptibility of net-baryon number in the proximity of the critical point for fixed $M=M_{CP}$ and for $T<T_{CP}$ (red squares) and $T>T_{CP}$ (green dots, results were scaled to improve readability of the figure) at the vanishing baryon chemical potential. Solid red and dashed green lines are the corresponding linear fits \ref{eq:linear_fit_Tm} and \ref{eq:linear_fit_Tp}, respectively.}
    \label{fig:scal_t_mu_0}
\end{figure}

In Fig.~\ref{fig:scal_t_mu_0} we show $\chi_4^B$ for $\mu_B=0$. Red squares correspond to $T<T_{CP}$ and the green dots to $T>T_{CP}$ (for the latter the results were multiplied by $5$ to improve the readability of the figure). The red solid and green dashed lines show the corresponding linear fits. We find $\varepsilon^-_4\approx0.665$ and $\varepsilon^+_4\approx0.668$, consistently with the MF prediction \eqref{eq:chi4_mu_0}. 

\begin{figure*}
    \centering
    \includegraphics[width=0.49\linewidth]{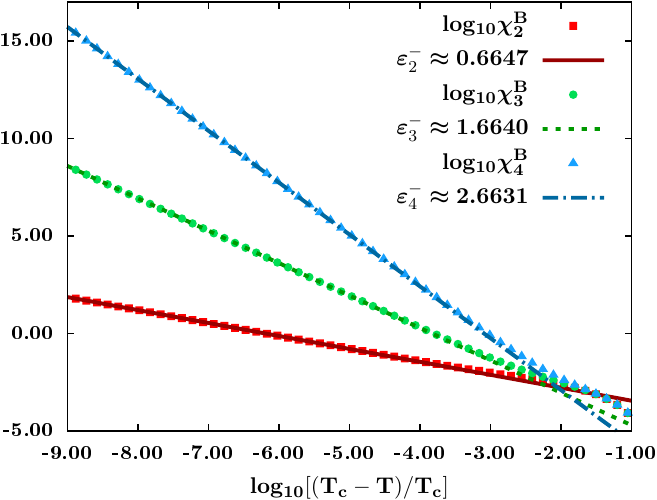}
    \includegraphics[width=0.49\linewidth]{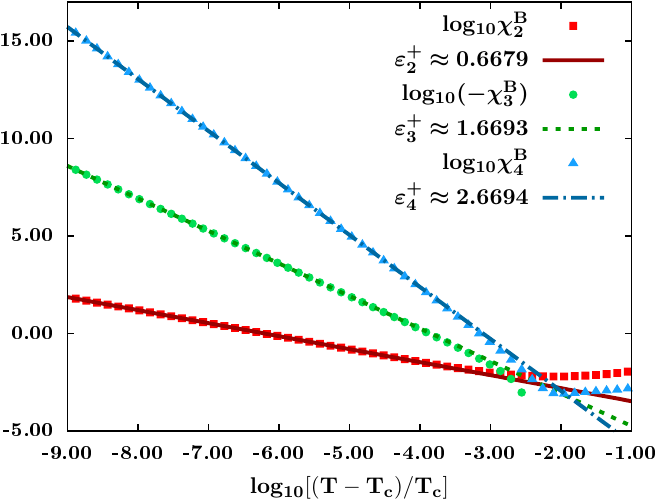}
    \caption{2nd (red squares), 3rd (green dots) and 4th (blue triangles) order susceptibilities of net-baryon number  in the proximity of the critical point for fixed $M=M_{CP}$ and for $T<T_{CP}$ (left panel) and $T>T_{CP}$ (right panel). Solid red, dashed green and dash-dotted blue lines show the corresponding linear fits \ref{eq:linear_fit_Tm} and \ref{eq:linear_fit_Tp}, respectively.}
    \label{fig:scal_t_mu}
\end{figure*}

Next, we discuss $\mu_B\neq0$ case. In Fig.~\ref{fig:scal_t_mu} we show $\chi_2^B$ (red squares), $\chi_3^B$ (green dots), and $\chi_4^B$ (blue triangles) for $T<T_c$ (left panel) and $T>T_c$ (right panel). The red solid, green dashed and blue dash-dotted lines show the corresponding linear fits. For $\chi_2^B$ we find $\varepsilon^-_2\approx0.665$ and $\varepsilon^+_2\approx0.668$, for $\chi_3^B$ $\varepsilon_3^-\approx1.664$ and $\varepsilon^+_3\approx1.669$ (where we multiplied $\chi_3^B$ by $-$ sign to keep the logarithm well-defined), and for $\chi_4^B$ we obtain $\varepsilon^-_4\approx2.663$ and $\varepsilon^+_4\approx2.669$. These values are consistent with predictions \ref{eq:chi2_mu}, \ref{eq:chi3_mu} and \ref{eq:chi4_mu}, respectively.

\begin{figure}
    \centering
    \includegraphics[width=\linewidth]{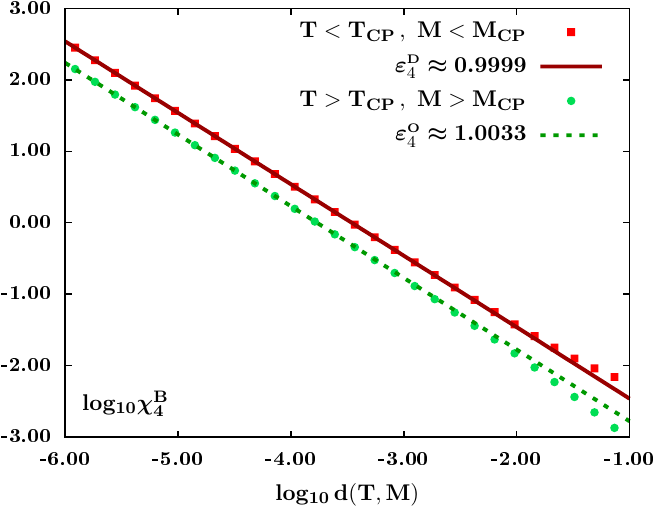}
    \caption{The 4th order susceptibility of the net-baryon number calculated along the tangent to the critical line when the CP is approached from the disordered (red squares) and ordered (green dots) phases. Solid red and dashed green lines show the corresponding linear fits.}
    \label{fig:scal_d_mu_0}
\end{figure}

\begin{figure*}
    \centering
    \includegraphics[width=0.49\linewidth]{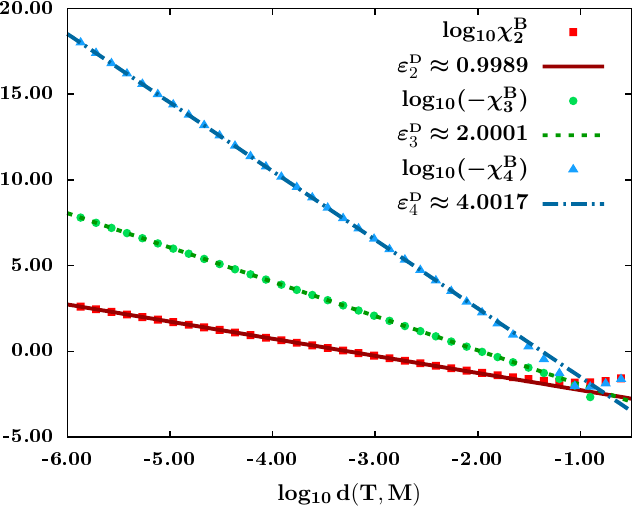}
    \includegraphics[width=0.49\linewidth]{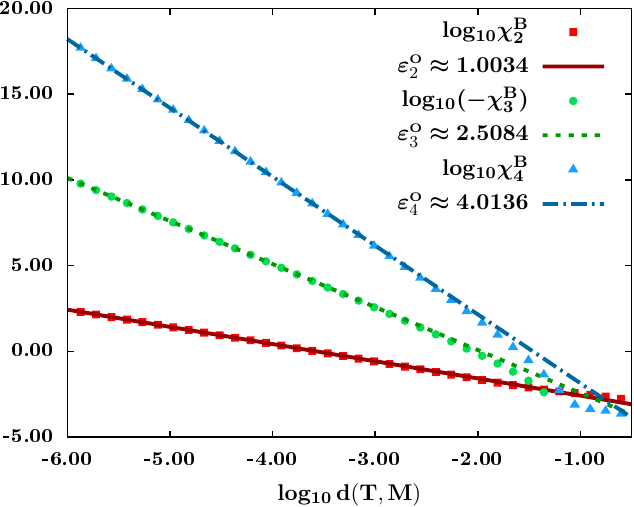}
    \caption{
    2nd (red squares), 3rd (green dots) and 4th (blue triangles) order susceptibilities of net-baryon number  in the proximity of the critical point, calculated along the tangent to the critical line when the CP is approached from the disordered (left panel) and ordered (right panel) phases. Solid red, dashed green and dash-dotted blue lines show the corresponding linear fits.}
    \label{fig:scal_d_mu}
\end{figure*}

Next, we investigate the critical exponents along the tangent to the critical line, which is parametrized as
\begin{equation}
    T=\alpha_{CP}(M-M_{CP})+T_{CP}\,,
\end{equation}
where the slope, $\alpha_{CP}$, is determined from Eq.~\eqref{eq:cp_slope} in Appendix~\ref{sec:CP_determination}. 
The tangents are marked by red dashes in Figs.~\ref{fig:chi4_3D} and~\ref{fig:chi3_3D}.

We quantify the distance to the CP as
\begin{equation}
    d(T,M)=\sqrt{(T/T_{CP}-1)^2+(M/M_{CP}-1)^2}
\end{equation}
and extract the critical exponents by performing the linear fit,
\begin{equation}
    \log_{10}\,\chi_n^B=-\varepsilon_{n}\,\log_{10}\,d(T,M)+b_n\,.
    \label{eq:linear_fit_d}
\end{equation}
Since the distance to CP defined in this manner is always positive, we denote the critical exponent obtained by approaching the CP from the disordered phase ($T<T_{CP}$ and $M<M_{CP}$) by $\varepsilon_n^{D}$ and the one obtained by approaching the CP from the ordered phase ($T>T_{CP}$ and $M>M_{CP}$) by $\varepsilon_n^{O}$.

For $\mu_B=0$, the fourth order cumulant is shown in Fig.~\ref{fig:scal_d_mu_0}. Red squares correspond to the approach from the direction of the disordered phase and green dots to the approach from the ordered phase side. The red solid and green dashed lines show the corresponding linear fits to Eq.~\eqref{eq:linear_fit_d}. We find $\varepsilon_4^{D}\approx0.9999$ and $\varepsilon^{O}_4\approx1.0033$. As previously, these predictions are consistent with the MF values \eqref{eq:chi4_mu_0}. 

For finite chemical potential ($\mu_B=1.5\,$GeV) the results are shown in Fig.~\ref{fig:scal_d_mu} where red squares correspond to $\chi_2^B$, green dots to $\chi_3^B$,  and blue triangles to $\chi_4^B$. The left panel corresponds to the approach from the disordered phase direction and the right panel to the approach from the ordered phase. Where necessary, we multiplied the corresponding susceptibilities by $-1$ to keep the logarithm well-defined. The red solid, green dashed and blue dash-dotted lines show the corresponding linear fits to Eq.~\eqref{eq:linear_fit_d}. For $\chi_2^B$ we find $\varepsilon_2^{D}\approx0.9989$ and $\varepsilon^{O}_2\approx 1.0034$, in case of $\chi_3^B$ we obtain $\varepsilon_3^{D}\approx 2.0001$ and $\varepsilon_3^{O}\approx 2.5084$, and for $\chi_4^B$ the critical exponents read $\varepsilon_4^{D}\approx 4.0017$ and $\varepsilon_4^{O}\approx 4.0136$. These values are consistent with predictions \ref{eq:chi2_mu}, \ref{eq:chi3_mu} and \ref{eq:chi4_mu}. In particular, we observe different critical exponent for $\chi_3^B$, depending on the direction of the approach to the CP. We summarized results on the MF criticality in Tab.~\ref{tab:mf_crit_exponents}.

\begin{table*}[]
\centering
\resizebox{\linewidth}{!}{
\begin{tabular}{ccccccccccccccccc}
\hline
 &
  \multicolumn{8}{c}{$\mu_B=0$} &
  \multicolumn{8}{c}{$\mu_B\neq0$} \\ \hline
Path &
  \multicolumn{4}{c}{$M=M_{CP}$} &
  \multicolumn{4}{c}{Tangent} &
  \multicolumn{4}{c}{$M=M_{CP}$} &
  \multicolumn{4}{c}{Tangent} \\ \hline
Direction &
  \multicolumn{2}{c}{$T<T_c$} &
  \multicolumn{2}{c}{$T>T_c$} &
  \multicolumn{2}{c}{Disordered phase} &
  \multicolumn{2}{c}{Ordered phase} &
  \multicolumn{2}{c}{$T<T_c$} &
  \multicolumn{2}{c}{$T>T_c$} &
  \multicolumn{2}{c}{Disordered phase} &
  \multicolumn{2}{c}{Ordered phase} \\ \hline
 &
  prediction &
  fit &
  prediction &
  fit &
  prediction &
  fit &
  prediction &
  fit &
  prediction &
  fit &
  prediction &
  fit &
  prediction &
  fit &
  prediction &
  fit \\ \hline
\multicolumn{1}{l}{$\chi_2^B$} &
  finite &
  -- &
  finite &
  -- &
  finite &
  -- &
  finite &
  -- &
  2/3 &
  0.6647 &
  2/3 &
  0.6679 &
  1 &
  0.9989 &
  1 &
  1.0034 \\
\multicolumn{1}{l}{$\chi_3^B$} &
  0 &
  -- &
  0 &
  -- &
  0 &
  -- &
  0 &
  -- &
  5/3 &
  1.6640 &
  5/3 &
  1.6693 &
  2 &
  2.0001 &
  5/2 &
  2.5084 \\
\multicolumn{1}{l}{$\chi_4^B$} &
  2/3 &
  0.6647 &
  2/3 &
  0.6684 &
  1 &
  0.9999 &
  1 &
  1.0033 &
  8/3 &
  2.6631 &
  8/3 &
  2.6694 &
  4 &
  4.0017 &
  4 &
  4.0136 \\ \hline
\end{tabular}}
\caption{Summary of the critical exponents of baryon number susceptibilities obtained in the MF approximation.}
\label{tab:mf_crit_exponents}
\end{table*}

\subsection{Critical exponents beyond the MF approximation}
\label{sec:beyond_mf}
To investigate critical properties of the net-baryon number fluctuations near the deconfinement critical point beyond the mean field approximation, we follow the scaling function approach. We assume that close to the deconfinement CP, the free energy can be written as~\cite{Ejiri:2005wq,Karsch:2023pga}
\begin{equation}
    f(T,M,\mu_B)=f_{r}(T,M,\mu_B)+f_{s}(T,M,\mu_B)\,,
\end{equation}
where $f_r$ is the regular, non-critical contribution, and $f_s$ is the singular part. The latter can be written in terms of the scaling function~\cite{Karsch:2023pga} which we parametrize as
\begin{equation}
    f_s(t,h)=t^{2-\alpha}\psi\bigg(\frac{h}{t^{2-\alpha-\beta}}\bigg)\,,
    \label{eq:scaling_fun}
\end{equation}
where $\alpha\approx0.11008$ and $\beta\approx0.32641$ are the 3D Ising model critical exponents\footnote{These were determined from the scaling dimensions $\Delta_\sigma=1/2+\eta/2\approx0.5181149$ and $\Delta_\varepsilon=3-1/\nu\approx1.41263$ calculated using the conformal bootstrap~\cite{Chang:2024whx}}, and $t$ and $h$ are the reduced temperature and magnetic field, for which we employ the same parametrization as in the MF case, Eqs.~\eqref{eq:h_t_param}. The critical exponents are obtained by taking the subsequent $\mub$-derivatives of the scaling function and extracting the most singular terms. We observe that the qualitative behavior of susceptibilities is the same as in the MF approximation but the divergences are stronger. Particularly, we find that at $\mu_B=0$ $\chi^B_2$ remains finite while $\chi_4^B$ diverges as $\gamma$. At finite density we predict that $\chi^B_2$ diverges as $\gamma$, $\chi^B_3$ as $\beta+2\gamma\approx2.8006$ and $\chi_4^B$ as $2\beta+3\gamma\approx4.3641$. These results are summarized in Tab.~\ref{tab:crit_exponents_beyond_MF}.

\begin{table}[]
\centering
\begin{tabular}{@{}ccc@{}}
\toprule
           & $\mu_B=0$ & $\mu_B\neq0$     \\ \midrule
$\chi_2^B$ & finite    & $\gamma\approx1.2371$         \\
$\chi_3^B$ & --         & $2\gamma+\beta\approx2.8006$  \\
$\chi_4^B$ & $\gamma\approx1.2371$  & $3\gamma+2\beta\approx4.3641$\\
\bottomrule
\end{tabular}
\caption{Prediction on critical exponents obtained from the scaling function~\eqref{eq:scaling_fun}.}
\label{tab:crit_exponents_beyond_MF}
\end{table}

\section{Estimates of the size of critical Region}
\label{sec:crit_region}

\begin{figure*}
    \centering
    \includegraphics[width=0.49\linewidth]{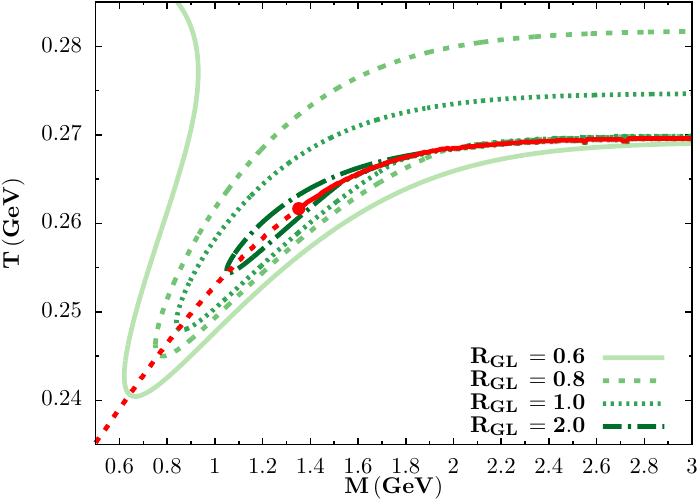}
    \includegraphics[width=0.49\linewidth]{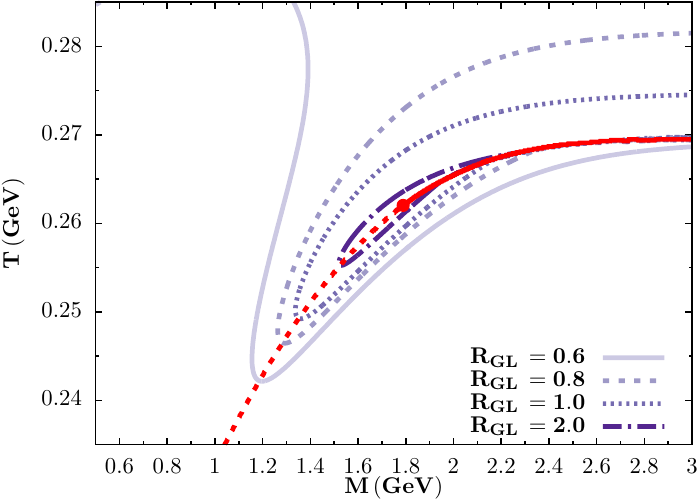}
    \caption{Estimate of critical region based on GL ratio for  $\mu_B=0$ (left) and $\mu_B=1.5\,$GeV (right). Solid red lines correspond to the first-order phase transition and dashed ones to the crossover.}
    \label{fig:crit_region} 
\end{figure*}

\begin{figure}
    \centering
    \includegraphics[width=\linewidth]{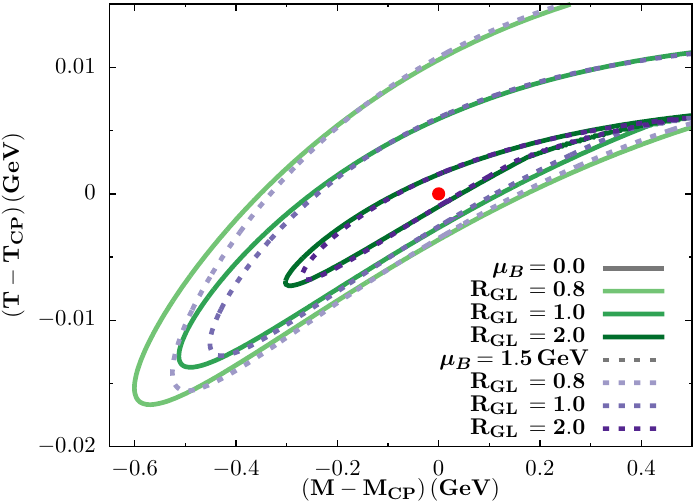}
    \caption{Contours of the GL ratio for $\mu_B=0$ (solid lines) and $\mu_B=1.5\,$GeV (dashed lines) relative to the corresponding critical point (red dot). }
    \label{fig:crit_region2}
\end{figure}

In this section we discuss the estimation of the size of the critical region within the current model. The Ginzburg–Landau (GL) criterion is based on the ratio
\begin{align}
R_{\rm GL}
= \frac{\langle (\Delta \phi)^2 \rangle}{\bar{\phi}^2},
\label{eq:GL01}
\end{align}
where $\phi$ denotes the order-parameter field and $\bar{\phi}$ its mean value.
When $R_{\rm GL} \gg 1$, fluctuations dominate over the mean field and the MF approximation ceases to be reliable. In standard textbook treatments~\cite{Goldenfeld:1992qy}, this ratio is primarily used to estimate the upper critical dimension. In Ref.~\cite{Lo:2026xuc}, however, it is repurposed as a quantitative diagnostic of the size of the critical region.

In this work, we compute $R_{\rm GL}$ for the present model to assess how robustly it characterizes the extent of the fluctuation-dominated regime. As a measure of the fluctuations of the order parameter we consider the susceptibility of the real part of the Polyakov loop, $\chi_L$ (see Eq.~\eqref{eq:chi_T_L}). The center symmetry is broken at high temperature, in contrast to a typical pattern of spontaneous symmetry breaking in which the ordered phase occurs below the critical temperature. This is taken into account by setting $\phi=1-\ell_R$. Thus, the $R_{GL}$ in the present model can be written as
\begin{equation}
    R_{GL}=\frac{\chi_R}{(1-\ell_R)^2}\,.
\end{equation}

We emphasize that the size of the critical region is not fixed by universality alone. While critical exponents are universal, the extent of the region in which non-MF behavior is visible depends on non-universal details, including regular contributions to the free energy. In this sense, the size of the critical region may be more directly relevant for phenomenology than the values of the critical exponents themselves. A reliable estimate of this region provides a realistic assessment of the range of validity of effective modeling.

We evaluate the GL ratio for phase diagrams at vanishing and finite baryon chemical potential. Figure~\ref{fig:crit_region} shows $R_{\rm GL}$ within the MF model considered here. The ratio provides a systematic characterization of the fluctuation-dominated region. Although the absolute size of the critical region depends on the operational definition of what constitutes a “large” ratio (here taken to be of order unity), meaningful comparisons can be made using a fixed criterion. Applying the same threshold at different chemical potentials, we find that the fluctuation-dominated region shrinks at finite $\mu_B$. This can be seen in Fig.~\ref{fig:crit_region2}, where contours of constant value of the GL ratio ($R_{GL}=0.8,$ 1 and 2) have been plotted relative to the corresponding critical points at zero (solid lines) and non-zero (dashed lines) baryon chemical potential. 

If one accepts that the GL criterion provides a reasonable proxy for the size of the critical region, this reduction at finite density may partly explain the practical difficulty of observing critical behavior. It underscores the importance of quantitatively controlling non-critical contributions. For effective models, this observation challenges the common argument that agreement at the level of universality class alone suffices for phenomenological relevance.

\section{Conclusions}
\label{sec:conclusions} 

In this work we examined the critical behavior of the first four cumulants of the net-baryon number in the vicinity of the deconfinement critical point with an effective Polyakov loop model. Using the Landau theory we derived the critical exponents in the MF approximation and confirmed our predictions with the numerical calculations performed using the effective model. 

At the vanishing baryon chemical potential $\chi_2^B$ remains finite while $\chi_4^B$ diverges. At the finite density, this patter changes as both the second-order susceptibility as well as the higher-order ones diverge. Divergences become stronger as the order of the susceptibility increases, in accordance with general expectations regarding the critical behavior of the net-baryon number susceptibilities~\cite{Stephanov:2008qz}. The same pattern is expected when fluctuations are included, as suggested by the scaling-function approach. The critical exponents obtained in this case are larger than the ones in the MF approximation. We also explored the critical region using the Ginzburg criterion and found that it shrinks with increasing baryon chemical potential.

Our findings suggests that net-baryon number susceptibilities are useful probes of criticality not only for physical quark masses, but also in the limit of heavy quarks and may provide an additional information on top on the conventionally used observables. It would be therefore worthwhile to investigate these quantities in a non-perturbative setting, for example using LQCD methods.

The estimate of the critical region presented here is based on a Landau MF approximation and therefore neglects spatial fluctuations of the order parameter. As emphasized in Ref.~\cite{Lo:2026xuc}, omission of the kinetic term is not a benign approximation: it can lead to misleading conclusions already at the level of locating the critical point and assessing its surrounding region. The present calculation provides a concrete realization of such a MF analysis in heavy‑quark QCD, but simultaneously illustrates its intrinsic limitations.
A faithful discussion of the critical region requires control over spatial correlations, which are encoded in the kinetic term of the effective action. For the Polyakov loop this term is difficult to construct due to its composite nature~\cite{Dumitru:2000in}, which largely explains its absence in most effective treatments. See Ref.~\cite{Hua:2026lcm} for a recent attempt. Our results, together with the observations of Ref.~\cite{Lo:2026xuc}, indicate that neglecting this contribution can qualitatively distort the interpretation of critical behavior in heavy‑quark QCD. Addressing this issue in a controlled effective description remains an important open problem and will be revisited in the future.

\begin{acknowledgments}
M. S. acknowledges the financial support of the Polish National Science Centre (NCN) under the Preludium grant 2020/37/N/ST2/00367 and through the program Excellence Initiative–Research University of the University of Wroclaw of the Ministry of Education and Science.
The work of P.~M.~Lo, K.~R. and C.~S. is supported by the National
Science Centre (NCN), Poland, under OPUS Grant
No. 2022/45/B/ST2/01527. K.~R. acknowledges the support
of the Polish Ministry of Science and Higher Education.
C.~S. acknowledges the support of the World
Premier International Research Center Initiative (WPI)
under MEXT, Japan.

\end{acknowledgments}

\appendix

\section{Explicit form of the MF net-baryon number cumulants}
\label{sec:app_chiB}
In this appendix we present the main steps of derivation of the explicit formulas for the net-baryon number cumulants\footnote{We follow the strategy outlined in Ref.~\cite{Fukushima:2017csk}}. We consider a general case of $P=P(T,\mu_B,\vec{\phi}(T,\mu))$,  where $\vec{\phi}(T,\mu)$ is the vector of order parameters, satisfying a set of gap equations,
\begin{equation}
\frac{\pt p}{\pt\phi_i}=0\,.
\label{eq:general_gap}
\end{equation}
Because of the $\mub$-dependence of the order parameters, the $\mub$-derivative acts as a total derivative. Thus, for the first derivative of the pressure we find
\begin{equation}
    \frac{\pt p}{\pt\mub}=p^{(1)}+\frac{\pt p}{\pt \phi_i}\frac{\pt \phi_i}{\pt\mub}=p^{(1)}\,,
\end{equation}
where the second term vanishes because of Eq.~\eqref{eq:general_gap} which holds for any baryon chemical potential. We also introduce a shorthand notation 
\begin{equation}
p^{(n)}=\frac{\partial^n (P/T^4)}{\partial (\mub)^n}\bigg\vert_{T,\,\vec{\phi}=const.\,}\,,    
\end{equation}
i.e. the $\mu_B$-derivative of the pressure is taken at the constant value of mean-fields.  

The second derivative reads 
\begin{equation}
    \frac{\pt^2 p}{\pt \mub^2}=\frac{\pt p^{(1)}}{\pt\mub}=p^{(2)}+\frac{\pt p^{(1)}}{\pt \phi_i}\frac{\pt \phi_i}{\pt\mub}\,.
\end{equation}
To evaluate  $\pt \phi_i/\pt\mub$ we take $\mub$-derivative of the gap equations \eqref{eq:general_gap},
\begin{eqnarray}
    \frac{\pt}{\pt\mub}\frac{\pt p}{\pt\phi_i}&=&\frac{\pt p^{(1)}}{\pt\phi_i}+\frac{\pt^2 p}{\pt\phi_i\pt\phi_j}\frac{\pt\phi_j}{\pt\mub}\nonumber\\
&=&\frac{\pt p^{(1)}}{\pt\phi_i}-\mathcal{C}_{ij}\frac{\pt\phi_j}{\pt\mub}=0,    
\end{eqnarray}
where $\mathcal{C}_{ij}=-\pt^2p/(\pt\phi_i\pt\phi_j)$ is the curvature matrix. Hence,
\begin{equation}
    \frac{\pt\phi_i}{\pt\mub}=\chi_{ij}\frac{\pt p^{(1)}}{\pt\phi_j}
\end{equation}
and the second-order cumulant can be written as
\begin{eqnarray}
\label{eq:chi2B_ex}
\chi_2^B=p^{(2)}+\frac{\pt p^{(1)}}{\pt \phi_i}\ctld_{ij}\frac{\pt p^{(1)}}{\pt\phi_j}\,.
\end{eqnarray}
In order to calculate higher-order cumulants, the derivative of the susceptibility matrix has to be found. It can be calculated by differentiating both sides of the following relation with respect to $\mub$,
\begin{equation}
    \mathcal{C}_{ij}\chi_{jk}=\delta_{ik}
\end{equation}
which leads to 
\begin{equation}
    \frac{\pt \ctld_{ij}}{\pt\mub}=-\ctld_{ik}\frac{\pt \mathcal{C}_{kl}}{\pt\mub}\ctld_{lj}\,.
\end{equation}
After calculating the derivative of the curvature matrix this relation becomes
\begin{equation}
\dmu\ctld_{ij}=\ctld_{ik}\left(\frac{\pt^2 p^{(1)}}{\pt \phi_k\phi_l}+\frac{\pt^3 p}{\pt \phi_k\pt \phi_l\pt\phi_m}\ctld_{mn}\frac{\pt p^{(1)}}{\pt\phi_n}\right)\ctld_{lj}\,.
\end{equation}
By repeating these steps, higher order net-baryon number cumulants can be evaluated. Particularly,
\begin{eqnarray}
    \chi_3^B&=&p^{(3)}+3\ctld_{ij}\frac{\pt p^{(2)}}{\pt \phi_i}\frac{\pt p^{(1)}}{\pt \phi_j}+3\frac{\pt^2 p^{(1)}}{\pt\phi_i\pt\phi_k}\ctld_{ij}\ctld_{kl}\frac{\pt p^{(1)}}{\pt\phi_j}\frac{\pt p^{(1)}}{\pt\phi_l}\nonumber \\
&+&\frac{\pt^3 p}{\pt \phi_k \pt \phi_l \pt \phi_m }\ctld_{ki}\ctld_{lj}\ctld_{mn}\frac{\pt p^{(1)}}{\pt\phi_i}\frac{\pt p^{(1)}}{\pt\phi_j}\frac{\pt p^{(1)}}{\pt\phi_m}\nonumber\\
\label{eq:chi3B_ex}
\end{eqnarray}
and
\begin{widetext}
\begin{eqnarray}
    \chi_4^B &=& p^{(4)}+3\ctld_{ij}\frac{\pt p^{(2)}}{\pt\phi_i}\frac{\pt p^{(2)}}{\pt\phi_j}+4\ctld_{ij}\frac{\pt p^{(3)}}{\pt\phi_i}\frac{\pt p^{(1)}}{\pt\phi_j}
    +6\frac{\pt^2 p^{(2)}}{\pt\phi_i\pt\phi_k}\ctld_{ij}\ctld_{kl}\frac{\pt p^{(1)}}{\pt\phi_j}\frac{\pt p^{(1)}}{\pt\phi_l}+12\frac{\pt^2 p^{(1)}}{\pt\phi_i\pt\phi_k}\ctld_{ij}\ctld_{kl}\frac{\pt p^{(1)}}{\pt\phi_j}\frac{\pt p^{(2)}}{\pt\phi_l}\nonumber\\
&+&4\frac{\pt^3 p^{(1)}}{\pt \phi_i\pt \phi_k\pt \phi_m}\ctld_{ij}\ctld_{kl}\ctld_{mn}\frac{\pt p^{(1)}}{\pt\phi_j}\frac{\pt p^{(1)}}{\pt\phi_l}\frac{\pt p^{(1)}}{\pt\phi_n}+6\frac{\pt^3 p}{\pt \phi_i\pt \phi_k\pt \phi_m}\ctld_{ij}\ctld_{kl}\ctld_{mn}\frac{\pt p^{(1)}}{\pt\phi_j}\frac{\pt p^{(1)}}{\pt\phi_l}\frac{\pt p^{(2)}}{\pt\phi_n}\nonumber\\
&+&12\frac{\pt^2 p^{(1)}}{\pt \phi_i\pt \phi_k}\ctld_{kl}\frac{\pt^2 p^{(1)}}{\pt \phi_l\pt \phi_m}\ctld_{ij}\ctld_{mn}\frac{\pt p^{(1)}}{\pt\phi_j}\frac{\pt p^{(1)}}{\pt\phi_n}+\frac{\pt^4 p}{\pt \phi_i \pt \phi_j \pt \phi_k \pt \phi_l}\ctld_{ip}\ctld_{jq}\ctld_{kr}\ctld_{ls}\frac{\pt p^{(1)}}{\pt\phi_p}\frac{\pt p^{(1)}}{\pt\phi_q}\frac{\pt p^{(1)}}{\pt\phi_r}\frac{\pt p^{(1)}}{\pt\phi_s}  \nonumber \\
&+&12\frac{\pt^2 p^{(1)}}{\pt \phi_i\pt \phi_k}\ctld_{kl}\frac{\pt^3 p}{\pt \phi_l\pt \phi_m\pt \phi_r}\ctld_{ij}\ctld_{mn}\ctld_{rs}\frac{\pt p^{(1)}}{\pt\phi_j}\frac{\pt p^{(1)}}{\pt\phi_n}\frac{\pt p^{(1)}}{\pt\phi_s}\nonumber\\
&+&3\frac{\pt^3 p}{\pt\phi_k \pt\phi_l \pt\phi_m}\ctld_{mp}\frac{\pt^3 p}{\pt\phi_p \pt\phi_q \pt\phi_r}\ctld_{ki}\ctld_{lj}\ctld_{qn}\ctld_{rs}\frac{\pt p^{(1)}}{\pt\phi_i}\frac{\pt p^{(1)}}{\pt\phi_j}\frac{\pt p^{(1)}}{\pt\phi_n}\frac{\pt p^{(1)}}{\pt\phi_s}\,.\nonumber\\
\label{eq:chi4B_ex}
\end{eqnarray}
\end{widetext}

\section{Numerical determination of the critical point location}
\label{sec:CP_determination}

In this appendix we discuss how the location of the critical point can be obtained precisely in a class of models described by an effective Polyakov loop potential~\eqref{eq:pot_gen}. We assume that in the proximity of CP, $\ell$ and $\lbar$ can be approximated as
\begin{eqnarray}
\ell  &=& \ell_{CP}+\varphi\,,\nonumber\\
\lbar &=& \lbar_{CP}+x\varphi\,,\nonumber\\
\label{eq:l_lbar_near_cp}
\end{eqnarray}
where $\ell_{CP}$ and $\lbar_{CP}$ are values of the Polyakov loop and its conjugate at the CP, and $x$ is the proportionality constant which takes into account the difference between $\ell$ and $\lbar$ at finite $\mu_B$. Consequently, close to the CP the full effective potential is approximated as
\begin{equation}
\frac{U(\ell,\lbar)}{T^4}\approx u_0+\frac{1}{2}a \varphi^2+\frac{1}{3!}b\varphi^3+\frac{1}{4!}c\varphi^4-h\varphi\,.
\end{equation}
The coefficients read
\begin{align}
     a&=\frac{\pt^2 (U/T^4)}{\pt\ell^2}
   +2x\frac{\pt^2(U/T^4)}{\pt\ell\pt\lbar}
   +x^2\frac{\pt^2 (U/T^4)}{\pt\lbar^2}\,,\\
   b&=\frac{\pt^3(U/T^4)}{\pt\ell^3}+3x\frac{\pt^3(U/T^4)}{\pt\ell^2\pt\lbar}
+3x^2\frac{\pt^3(U/T^4)}{\pt\ell\pt\lbar^2}\nonumber\\
&+x^3\frac{\pt^3(U/T^4)}{\pt\lbar^3}\,,\\
c&=\frac{\pt^4(U/T^4)}{\pt\ell^4}+ 4x\frac{\pt^4(U/T^4)}{\pt\ell^3\pt\lbar}+6x^2\frac{\pt^4(U/T^4)}{\pt\ell^2\pt\lbar^2}\nonumber\\
&+4x^3\frac{\pt^4(U/T^4)}{\pt\ell\pt\lbar^3}+x^4\frac{\pt^4(U/T^4)}{\pt\lbar^4}  \\
h&=\frac{\pt(U/T^4)}{\pt\ell}+x\frac{\pt(U/T^4)}{\pt\lbar}\,.
\end{align}
These coefficients are functions of temperature, quark mass and baryon chemical potential. In this work, the latter is treated as an parameter, and thus for a given $\mu_B$ the values of $T_{CP}$, $M_{CP}$, $\ell_{CP}$, $\lbar_{CP}$ and $x_{cp}$ need to be determined. 

To formulate a closed set of equations we note that at the critical point the potential becomes flat, i.e. $a=b=0$. 
Two additional constraints arise from the gap equations~\eqref{eq:gap_eqs} (which sets $h=0$). The last equation can be obtained from the requirement that at the critical point, the inverse of the curvature matrix of the original potential becomes singular, i.e. 
\begin{equation}
    det(\mathcal{C})=\frac{\pt^2 (U/T^4)}{\pt\ell^2}\frac{\pt^2 (U/T^4)}{\pt\lbar^2}-\bigg(\frac{\pt^2 (U/T^4)}{\pt\ell\pt\lbar}\bigg)^2=0\,.
\end{equation}
Using these equations, $T_{CP}$, $M_{CP}$, $\ell_{CP}$, $\lbar_{CP}$ and $x_{CP}$ can be determined in function of $\mu_B$. For $\mu_B=0$, $N_f=2$ we find
\begin{eqnarray}
    T_{CP}^{\mu_B=0}&\approx&0.26167\,\text{GeV}\,,\nonumber\\
    M_{CP}^{\mu_B=0}&\approx&1.35160\,\text{GeV}\,,\nonumber\\
    \ell_{CP}^{\mu_B=0}=\lbar_{CP}^{\mu_B=0}&\approx&0.17332\,,\nonumber\\
    x_{CP}^{\mu_B=0}&=&1\,.\nonumber\\
    \label{eq:num_cp_res_mu_0}
\end{eqnarray}
Critical mass and temperature agree with Ref.~\cite{Lo:2014vba}. For $\mu_B=1.5\,$GeV $N_f=2$ we find
\begin{eqnarray}
    T_{CP}^{\mu_B=1.5\,\text{GeV}}&\approx&0.26208\,\text{GeV}\,,\nonumber\\
    M_{CP}^{\mu_B=1.5\,\text{GeV}}&\approx&1.78911\,\text{GeV}\,,\nonumber\\
    \ell_{CP}^{\mu_B=1.5\,\text{GeV}}&\approx&0.16384\,,\nonumber\\
    \lbar_{CP}^{\mu_B=1.5\,\text{GeV}}&\approx&0.18772\,,\nonumber\\
    x_{CP}^{\mu_B=1.5\,\text{GeV}}&\approx&0.87087\,.\nonumber\\
    \label{eq:num_cp_res_mu}
\end{eqnarray}

The slope of the critical line at the CP can be determined from the linear term,
\begin{equation}
    h=\frac{\pt(U/T^4)}{\pt\ell}+x\frac{\pt(U/T^4)}{\pt\lbar}.
\end{equation}
The solutions of $h(T,M)=h_0$ yield the lines of constant $h$. Particularly, the $h=0$ line is tangent to the line of the first-order phase transition at the CP and thus its slope can be obtained from
\begin{equation}
    \alpha_{CP}\equiv\frac{dT}{dM}\bigg\vert_{CP}=-\frac{\frac{\pt^2 (U/T^4)}{\pt M\pt\ell}+x\frac{\pt^2 (U/T^4)}{\pt M\pt\lbar}}{\frac{\pt^2(U/T^4)}{\pt T\pt\ell}+x\frac{\pt^2(U/T^4)}{\pt T\pt\lbar}}\bigg\vert_{CP}\,,
    \label{eq:cp_slope}
\end{equation}
where subscript CP denotes that derivatives are evaluated using the CP values of $T$, $M$, $\ell$, $\lbar$ and $x$. For $\mu_B=0$, $N_f=2$ we find
\begin{eqnarray}
    \alpha_{CP}^{\mu_B=0}&\approx&0.01993\,.
\end{eqnarray}
and for $\mu_B=1.5\,$GeV $N_f=2$
\begin{eqnarray}
    \alpha_{CP}^{\mu_B=1.5\,\text{GeV}}&\approx&0.02070\,.
\end{eqnarray}

\end{document}